\documentclass[pra,10pt,a4paper,twocolumn,notitlepage,superscriptaddress]{revtex4-1}

\usepackage{dsfont,amsmath,mathtools,bm,mathbbol,soul,diagbox,multirow}
\usepackage[normalem]{ulem}
\usepackage{xcolor}
\usepackage{hyperref}
\hypersetup{
    colorlinks,
    citecolor=black,
    filecolor=black,
    linkcolor=[rgb]{0,.5,.8},
    urlcolor=blue
}
\usepackage[utf8]{inputenc}
\usepackage[T1]{fontenc}

\newcommand{\bra}[1]{\langle #1 |}
\newcommand{\ket}[1]{| #1 \rangle}

\begin{document}	

\title{Majorization-based benchmark of the complexity of quantum processors}

\author{Alexandre B. Tacla}
\author{Nina Machado O'Neill}
\affiliation{Centro Brasileiro de Pesquisas F\'{\i}sicas,
Rua Dr. Xavier Sigaud, 150, Rio de Janeiro, RJ, Brazil}

\author{Gabriel G. Carlo}
\affiliation{Departamento de F\'{\i}sica, Comisi\'on Nacional de Energ\'\i a At\'omica, Avenida del Libertador 8250, (1429) Buenos Aires, Argentina}

\author{Fernando de Melo}
\author{Ra\'ul O. Vallejos}
\affiliation{Centro Brasileiro de Pesquisas F\'{\i}sicas,
Rua Dr. Xavier Sigaud, 150, Rio de Janeiro, RJ, Brazil}

\begin{abstract}
	Here we investigate the use of the majorization-based indicator introduced in [R. O. Vallejos, F. de Melo, and G. G. Carlo, Phys. Rev. A 104, 012602 (2021)] as a way to benchmark the complexity within reach of quantum processors. By considering specific architectures and native gate sets of currently available technologies, we numerically simulate and characterize the operation of various quantum processors. We characterize their complexity for different native gate sets, qubit connectivity and increasing number of gates. We identify and assess quantum complexity by comparing the performance of each device against benchmark lines provided by randomized Clifford circuits and Haar-random pure states. In this way, we are able to specify, for each specific processor, the number of native quantum gates which are necessary, on average, for achieving those levels of complexity. Lastly, we study the performance of the majorization-based characterization in the presence of distinct types of noise. We find that the majorization-based benchmark holds as long as the circuits' output states have, on average, high purity ($\gtrsim 0.9$). In such cases, the indicator showed no significant differences from the noiseless case.
\end{abstract}

\maketitle

\section{Introduction} 
\label{sec:introduction}

Various benchmarking techniques have been proposed over the recent years to evaluate and calibrate the operation and performance of quantum gates~\cite{Knill08,Magesan12,Magesan12b,Dugas15,Cross2016} and of quantum processors 
\cite{Emerson07,Magesan11,Cross2019,Wack2021,Wang2022}. A desirable benchmarking technique should be ideally practical to implement, architecture independent, and scale favourably with increasing number of qubits. In order to be applicable to the current generation of noisy intermediate-scale quantum (NISQ) processors~\cite{Preskill2018}, it should also be capable of reliably identifying the potential of obtaining quantum computational advantage despite the presence of noise. Ultimately, these methods form a toolbox of metrics and tests that help validate noisy quantum computations and quantify the performance of today's quantum computers. Examples of popular techniques that are currently in use include quantum volume~\cite{Bishop17,Cross2016, Cross2019, Moll18, Baldwin2022}, randomized benchmarking~\cite{Knill08, Magesan11,Magesan12,Magesan12b,Brown18, Hashagen2018, Helsen2019,mckay2020,Helsen2022, Proctor2022,liu2022}, and cross-entropy benchmarking~\cite{Neill2018,Boixo2018}, which has been implemented in recent quantum-advantage experiments~\cite{Arute2019,Wu2021,Zhu2022}.

As shown recently in Ref.~\cite{vallejos01}, the complexity of random quantum circuits can be unveiled  using a majorization-based criterion, namely the fluctuations (standard deviation) of the Lorenz curves of the circuit output states. In particular, it was shown that this criterion can serve as a heuristic complexity indicator, capable of discriminating between universal and non-universal classes of random quantum circuits. 
Moreover, it was also able to correctly identify the complexity of some non-universal but not classically efficiently simulatable random quantum circuits. Sucessful applications in reservoir quantum computing were also 
recently reported \cite{Domingo2022}.

Here we investigate the use of this indicator, as a potential way to benchmark the complexity within reach of presently available quantum processors. We consider a simple architecture-independent protocol, which only requires random circuit samples (which do not need to be known) and the measurement of the probabilities in the computational basis. By numerically simulating the operation of several currently available universal, gate-model quantum processing units (QPUs), we characterize their complexity for various native gate sets, qubit connectivity and increasing number of gates. To identify quantum complexity, we analyze each case in relation to key reference curves calculated for randomized Clifford circuits and Haar-random pure states. This procedure allows us to specify, for each QPU, the number of native quantum gates which are necessary, on average, for reaching those levels of complexity. Finally, we study how these results change as noise is added to the QPU.  Essentially, we find that the majorization-based benchmark holds as long as the circuits' output states have, on average, high purity ($\gtrsim 0.9$). For those cases, the indicator showed no significant deviation from the noiseless case.

This article is organized as follows. In Sec.~\ref{sec:method}, we review the majorization-based criterion. In Sec.~\ref{sec:noiseless_quantum_processors}, we perform the majorization-based characterization of various noiseless quantum processors, including several IBM and Rigetti architechtures. In Sec.~\ref{sec:noisy_quantum_processors}, we apply the majorization-based indicator to noisy QPUs. We first consider, in Sec.\ref{sub:near_perfect_qpus}, the optimistic case of near-perfect QPUs. In this case, all native quantum gates are assumed perfect. However, idle qubits may experience noise in the form of amplitude damping or pure dephasing. Then, in Sec.~\ref{sub:faulty_qpus}, we study the more realistic case of faulty QPUs, whose imperfections we model as depolarizing channels. In this case, random Pauli errors may occur whenever a quantum gate is applied. Lastly, we give our conclusion and final remarks in Sec.~\ref{sec:discussion}.


\section{Method} 
\label{sec:method}

As we mentioned above, the majorization-based benchmarking procedure we consider here basically consists of computing the fluctuations of the Lorentz curves of the output probabilities of random quantum circuits for a given QPU. We briefly review below the key points of our characterization method. Further details can be found in Ref.~\cite{vallejos01}.

Majorization defines a partial ordering between two vectors, which establishes whether the components of one vector are more evenly distributed (disordered) than the components of the other~\cite{marshall09}. Given any two vectors $\textbf{p}, \textbf{q} \in \mathds{R}^N$, we say that \textbf{p} is majorized by \textbf{q} (or \textbf{q} majorizes \textbf{p}), denoted by $\textbf{p} \prec \textbf{q}$, if
\begin{align}
    \sum_{i=1}^{k} p_{i}^{\downarrow} & \leq \sum_{i=1}^{k} q_{i}^{\downarrow}, \quad 1\leq k < N, \label{eq:major_ineq}\\
    \sum_{i=1}^{N} p_{i} & = \sum_{i=1}^{N} q_{i}\label{eq:normalization}.
\end{align}
Here, the superscript $^{\downarrow}$ denotes that the vector components are sorted in non-increasing order. As we will be dealing exclusively with probability vectors, Eq.~(\ref{eq:normalization}) is trivially satisfied, whereas condition~(\ref{eq:major_ineq}) implies that probability distributions which are mostly concentrated in fewer components majorizes distributions that are spread out over the masurement basis set. Indeed, for any $p_i\ge 0$ and $\sum_{i=1}^N p_i =1$, it follows that
\begin{align*}
    (1/N, 1/N, \ldots, 1/N) \prec (p_1, p_2, \ldots, p_N) \prec (1, 0, \ldots, 0).
\end{align*}
It is convenient, from now on, to denote the partial sums in condition~(\ref{eq:major_ineq}) by $F_p(k)$ and $F_q(k)$, which we will refer to as the $k$-th cumulant of $\textbf{p}$ and $\textbf{q}$. Thus, we can restate the majorization condition as follows: if $\textbf{p} \prec \textbf{q}$, then $F_p(k) \leq F_q(k)$ for $1\leq k < N$. Equivalently, if we plot the cumulants $F_p(k)$ and $F_q(k)$ vs $k/N$, also known as \emph{Lorenz curves}, then \textbf{q} majorizes \textbf{p} iff the Lorenz curve for \textbf{q} is above the curve for \textbf{p} for all values of $k/N$.

Although the Lorenz curves allow one to decide whether a probability distribution is more disordered than another, it has been shown that they do not fully identify different complexities of random circuits (not even on average), having failed to differentiate universal from non-universal families of quantum circuits~\cite{vallejos01}. A correct identification, however, was achieved through the fluctuations of the Lorenz curves. Thus, as our general protocol, we consider an ensemble of $n$-qubit random quantum circuits $\{U\}$, by sampling uniformly from the set of quantum gates native to a given quantum processor. Unless explicitly stated, all random circuits start with all qubits in the $|0\ldots 0\rangle= |0\rangle^{\otimes n}$ state, and end with a measurement in the computational basis $\{|i\rangle\}$ -- with $i\in\{0,1\}^n$ a $n$-bit string. For each random circuit, we compute the output distribution, $p_U (i) = \left|\bra{0\ldots 0} U \ket{i}\right|^2$, sort it in descending order, and evaluate the cumulants $F_{p_U}(k)$ -- with $k\in\{1,\ldots,2^n\}$.  Sampling a sufficiently large number of quantum circuits (to be discussed below), we then evaluate the cumulant fluctuation as:
\begin{align}\label{eq:stdF}
    {\rm std}\,[F_{p_U} (k)] = \sqrt{\langle F_{p_U}^2 (k) \rangle - \langle F_{p_U}(k) \rangle^2},
\end{align}
where the averages are taken over the ensemble of random circuits -- obtained by sampling uniformly from the gate set. Each device is then characterized in terms of the fluctuations of the Lorenz curves for increasing numbers of quantum gates and for a sufficiently large number of random circuits. 

To identify quantum complexity and assess the potential to reach quantum advantage, we directly compare the characteristic fluctuation curve for each $n$-qubit device to key reference curves calculated for $(i)$ randomized $n$-qubit Clifford circuits (denoted from now on as Cliff-$n$) and $(ii)$ $n$-qubit Haar-random pure states (denoted from now on as Haar-$n$). The characteristic curve for Cliff-$n$ circuits provides a reference for identifying typical complexity of random quantum circuits that can be efficiently simulated by classical means (when $n$ is large). Specifically, Cliff-$n$ circuits are generated from the set 
\{CNOT, H, S\} (CNOT: Controlled-NOT, H: Hadamard, S: $\pi/2$-phase gate), starting from a random pure separable state and with no qubit connectivity constraint. For a sufficiently large number of gates, we have verified numerically that this procedure yields a limiting curve. The limiting Cliff-$n$ curve, however, does not impose a limit for $n$-qubit random circuits constructed from non-Clifford gate sets. On the other hand, the characteristic  Haar-$n$ curve does provide\sout{s} a lower limit (see results below) for universal (non-noisy) gate sets and, hence, serves as our reference for identifying quantum complexity that is beyond the reach of classical computations (when $n$ is large). 


\section{Noiseless quantum processors} 
\label{sec:noiseless_quantum_processors}

\begin{figure}[tbp]  
  \centering
    \includegraphics[width=.5\textwidth]{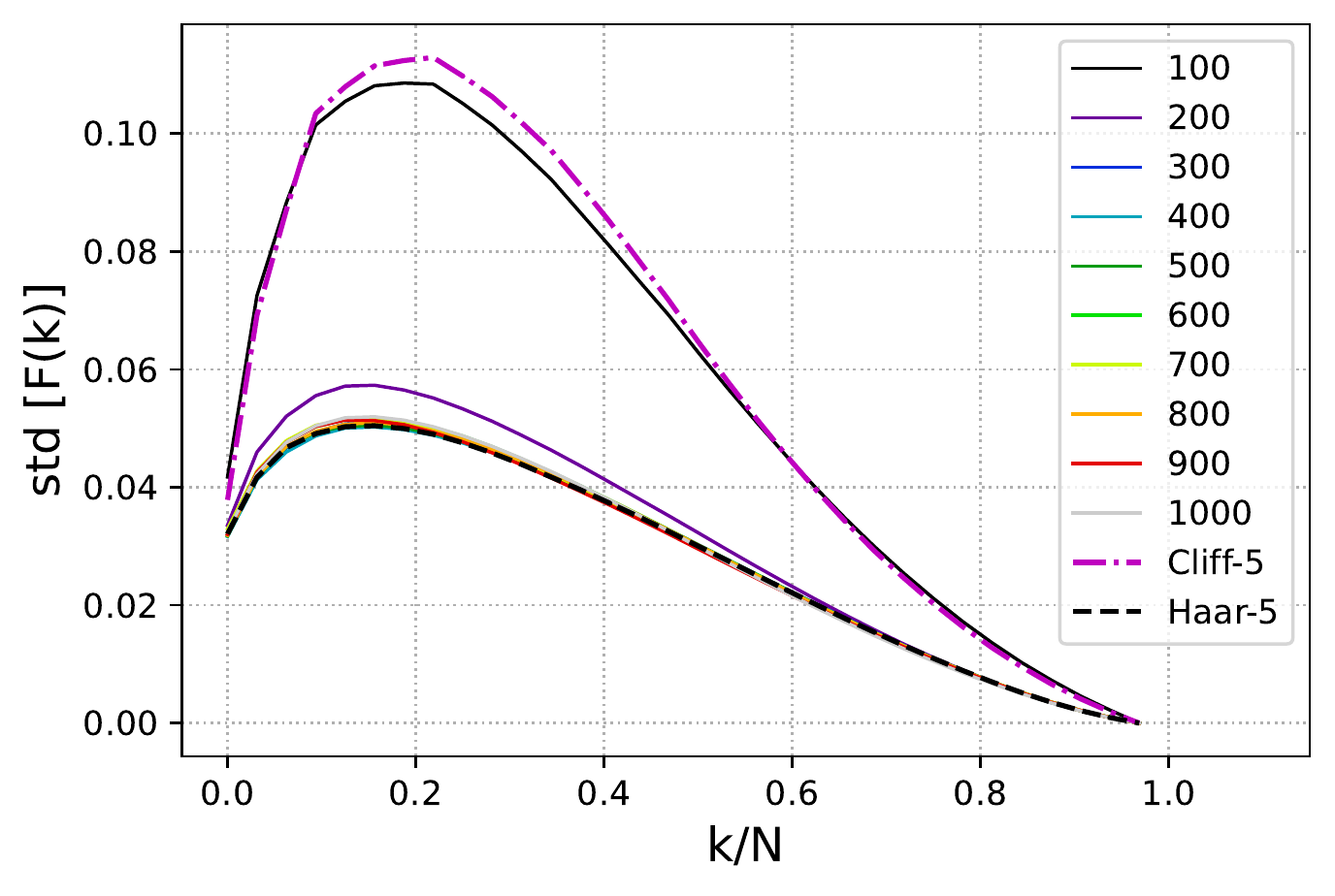}
  \caption{Majorization-based characterization of the five-qubit IBM Q Yorktown device with ``bow tie'' topology. Here we plot the fluctuations of the Lorenz curves for 5000 random circuits and increasing number of gates. The top dot-dashed curve is the result for Cliff-5, which is a limiting curve for 5-qubit random Clifford circuits. For circuits constructed from non-Clifford gate sets, Cliff-5 serves as a reference line, which suggests that, on average, this IBM device running random circuits with 100 gates has reached a similar level of complexity. However, as the number of gates increases and, hence, circuit complexity grows, we see that the corresponding fluctuation curves decrease toward Haar-5 (dashed line) and have no longer non-trivial coincidences with Cliff-5. For 300 gates or more, all fluctuation curves are visually indistinguishable from Haar-5, suggesting that they all have comparable levels of complexity. 
  }
  \label{fig:plots_stdF_Q5}
\end{figure}

In this section, we numerically simulate the operation of various QPU systems, including currently available architectures from IBM and Rigetti, and apply the majorization-based indicator~(\ref{eq:stdF}) to characterize their complexity. Here we assume all systems to be completely free of noise. We defer the study of noisy quantum processors to Sec.~\ref{sec:noisy_quantum_processors}. In all simulations presented below, we consider key specific parameters, including the device's native gate set, qubit connectivity, and number of qubits~\cite{kuatomu}. IBM quantum devices have the native gate set $G_{\rm IBM}$ =\{X, $\sqrt{X}$, RZ, CNOT\}, where RZ rotations can be of an arbitrary angle. Since Pauli X can be trivially generated by $\sqrt{X}$, we do not explicitly consider it in our simulations. Rigetti's quantum processors, on the other hand, can natively perform the gates from the set $G_{\rm Rig}$ = \{RX, RZ, CZ\}, where RZ rotations can be of an arbitrary angle, but RX rotations are restricted to $\pm \pi/2$ and $\pm \pi$ \cite{pyquil_documentation}. Both IBM and Rigetti processors can also natively measure in the computational basis. 

\begin{figure}[h]  
  \centering
    \includegraphics[width=.4\textwidth]{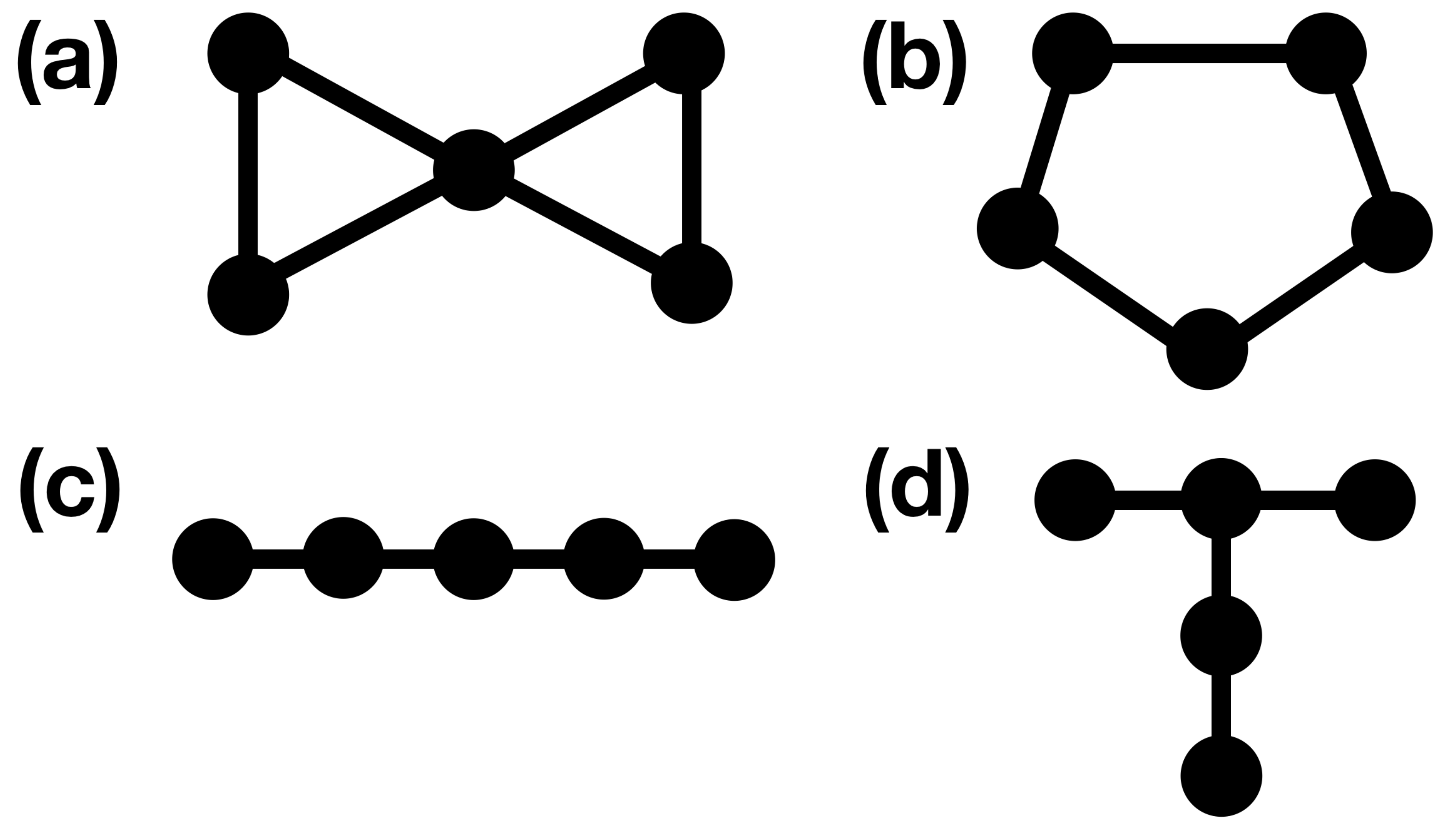}
  \caption{(a) IBM Q Yorktown (bow-tie) connectivity map, with an average number of qubit connections of $\bar n_c = 2.4$. Other examples we considered of 5-qubit QPU geometries~\cite{ibm_url}: (b) ring ($\bar n_c = 2$), (c) line (IBM Q Manila -- $\bar n_c = 1.6$), (d) ``T'' (IBM Q Belém -- $\bar n_c = 1.6$).}
  \label{fig:plots_IBM5_topologies}
\end{figure}

As a first example, we consider the early five-qubit IBM Q Yorktown device with a ``bow-tie'' qubit connectivity \cite{ibm_url}, as illustrated in Fig.~\ref{fig:plots_IBM5_topologies}(a). 
Figure~\ref{fig:plots_stdF_Q5} shows the characteristic fluctuations computed for an ensemble of 5000 random quantum circuits and for increasing number of quantum gates. The dot-dashed curve at the top is the asymptotic result for Cliff-5, which sets a limit for 5-qubit random Clifford circuits. The dashed curve at the bottom is the result for Haar-5, which sets a lower limit for the fluctuations of 5-qubit random quantum circuits. Thus, we can see that the fluctuations for 100 gates have a significant range of (visual) coincidence with Cliff-5. However, as the number of gates increases (and complexity grows), the fluctuations decrease and approach the Haar-5 curve. For 200 gates, fluctuations are already comparable to Haar-5; they have similar shape and height, and visually coincide for $k/N\gtrsim 0.6$. For 300 gates or more, all curves become visually indistinguishable from Haar-5. 

\begin{figure}[htb] 
  \centering
    \includegraphics[width=.5\textwidth]{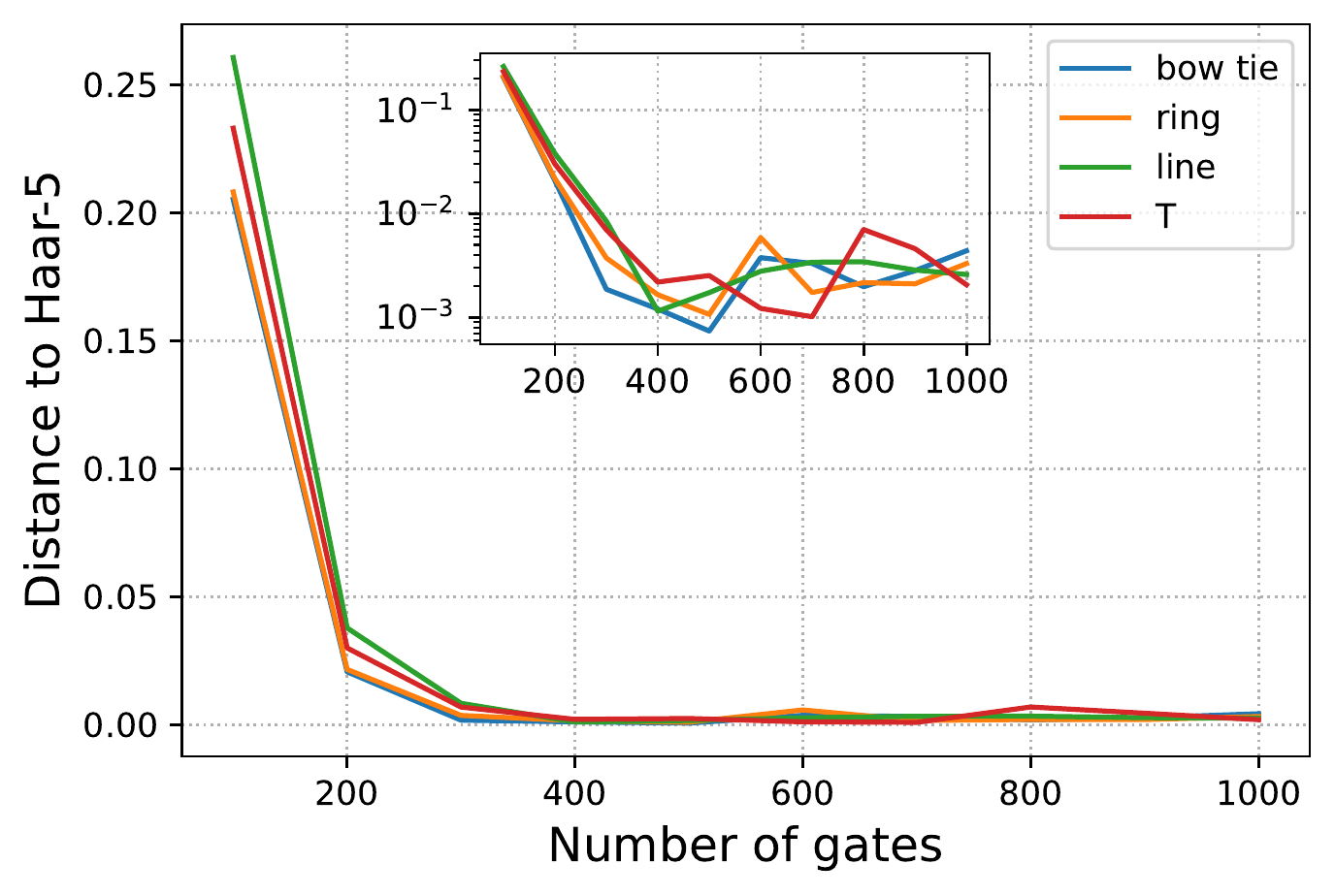}
  \caption{Distance to Haar-5, as defined in Eq.~(\ref{eq:DH}), as a function of the number of quantum gates for all 5-qubit geometries shown in Fig.~\ref{fig:plots_IBM5_topologies} with the IBM native gate set. The inset shows the same plot, but in logarithmic scale. The ``bow-tie'' and ``ring'' geometries have the quickest convergence to Haar-5, as they have the largest average qubit connectivity (2.4 and 2.0, respectively). Both ``line'' and ``T'' geometries have a lower average qubit connectivity (1.6 for both), and thus require a larger number of gates in order to coincide with Haar-5.}
  \label{fig:plots_distHaar_N5}
\end{figure}

Similar results can be seen for other 5-qubit IBM devices with different qubit connectivities. 
To see this more clearly, we compute the distance between each fluctuation curve 
(${\rm std}[F(k)]$) and its respective Haar-$n$ benchmark line (${\rm std}[F_H(k)]$), 
given by

\begin{equation}\label{eq:DH}
	D_{H} = \sqrt{\sum_{k=1}^{2^n}\Big({\rm std}[F(k)] - {\rm std}[F_{H}(k)]\Big)^2},
\end{equation}
for all qubit geometries illustrated in Fig.~\ref{fig:plots_IBM5_topologies}. 
Results are shown in Fig.~\ref{fig:plots_distHaar_N5}.

The ``bow-tie'' and ``ring'' geometries have the largest qubit connectivities we considered, with an average number of qubit connections of $\bar n_c = 2.4$ and $\bar n_c = 2$. Not surprisingly, for those geometries, convergence to Haar-5 is achieved with fewer gates than for both ``line'' and ``T'' geometries, which have a lower average qubit connectivity ($\bar n_c = 1.6$ for both). Nonetheless, all devices coincide with Haar-5 (within visual resolution) for 300 gates. 

\begin{figure}[htbp] 
  \centering
    \includegraphics[width=.45\textwidth]{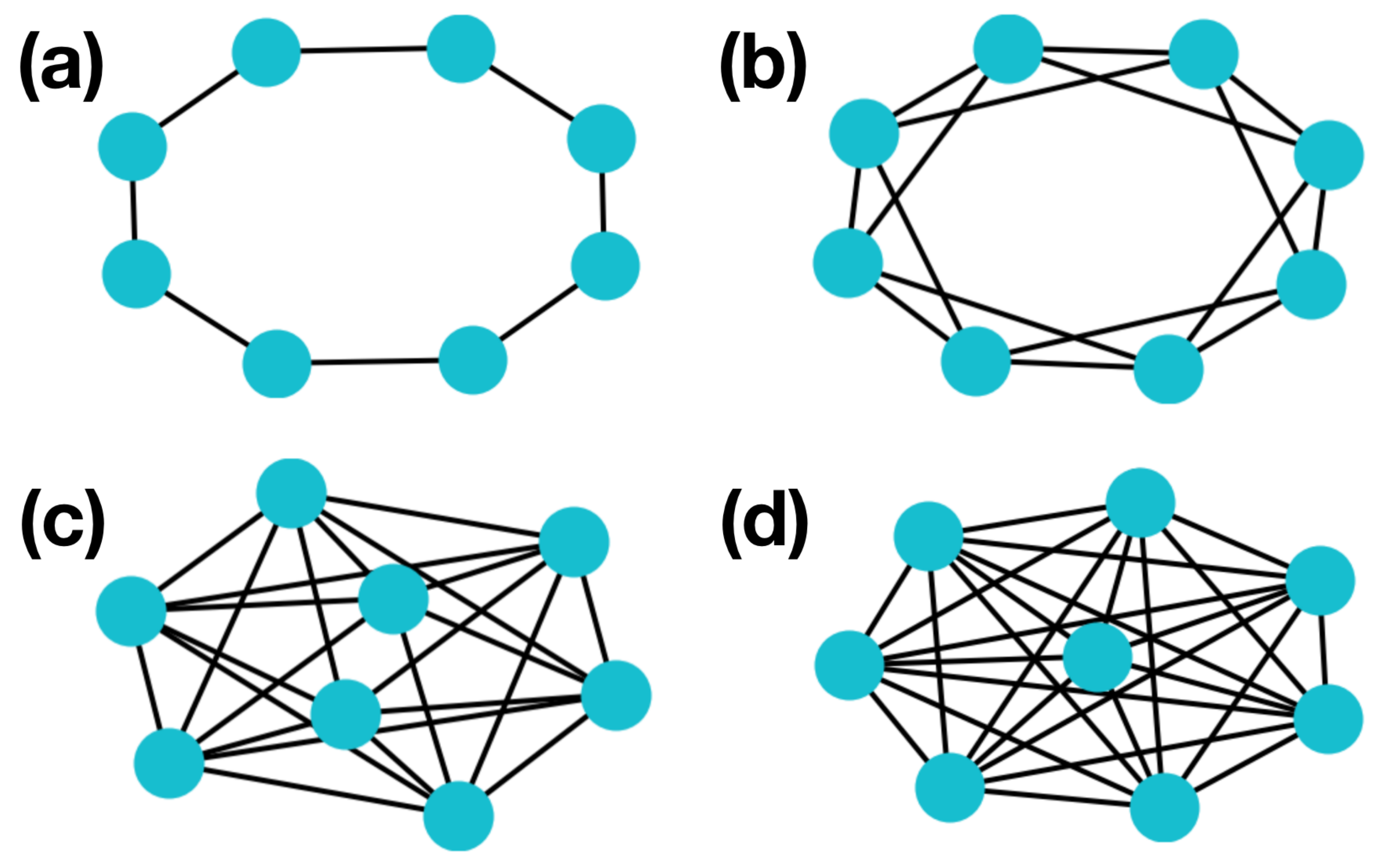}
  \caption{Connectivity maps for (a) Rigetti Agave 8-qubit, ``ring'' QPU, with nearest-neighbor connectivity ($n_c = 2$), and other examples we considered of 8-qubit QPU geometries with  increasing qubit connectivity: (b) $n_c = 4$, (c) $n_c = 6$, and (d) $n_c = 7$ (all-to-all).}
  \label{fig:plots_Rigetti_topologies}
\end{figure}

\begin{figure}[h] 
  \centering
    \includegraphics[width=.5\textwidth]{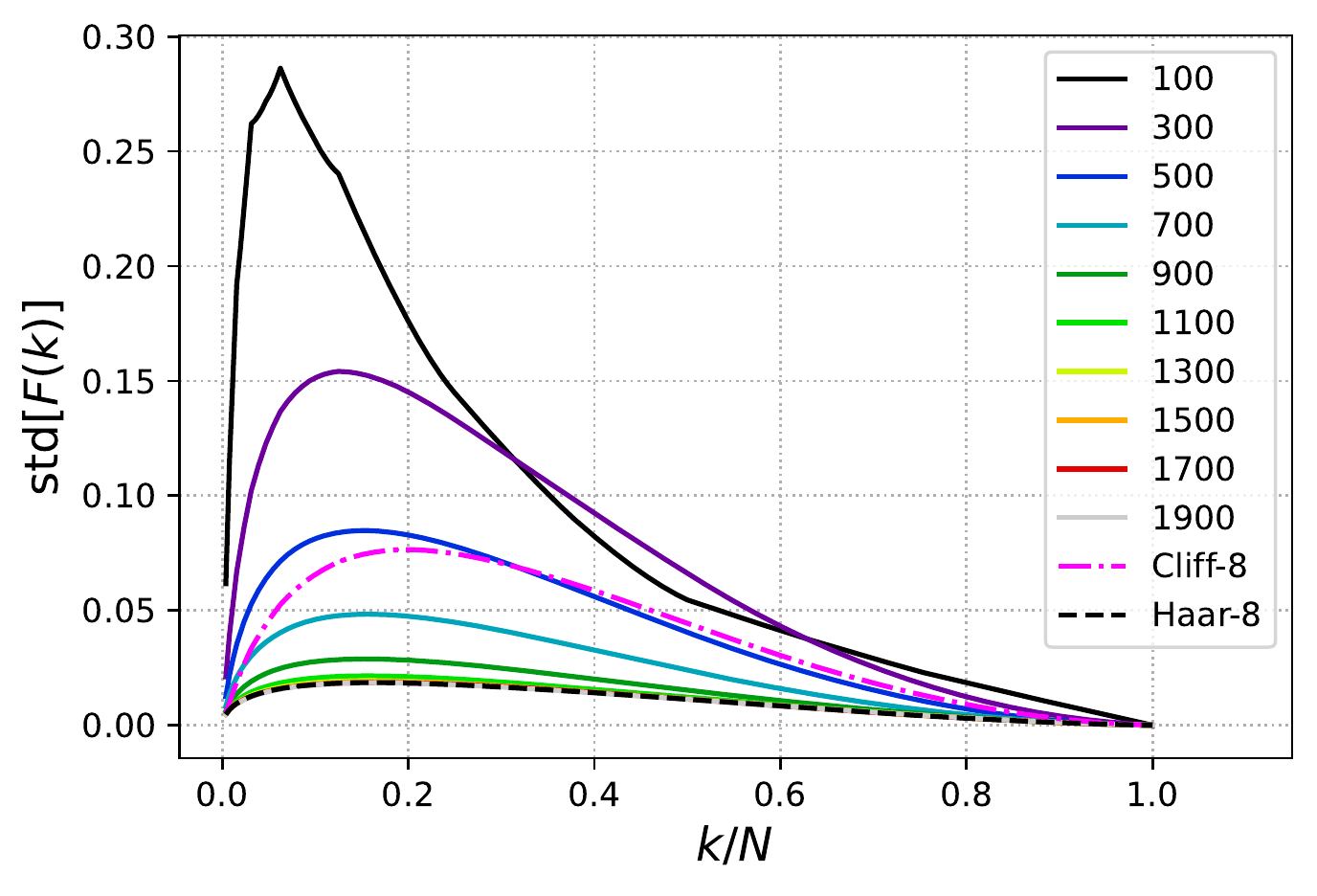}
  \caption{Majorization-based characterization of the Rigetti 8Q Agave device with ring topology and nearest-neighbor connectivity. Here we plot the fluctuations of the Lorenz curves for 5000 realizations and increasing number of gates. The dot-dashed curve is the result for Cliff-8. 
  For circuits constructed from non-Clifford gate sets, Cliff-8 serves as a guideline, which suggests that the Agave device running random circuits with 500 gates reaches, on average, a similar level of complexity. As the number of gates increases further, we see that the corresponding fluctuation curves decrease toward Haar-8 (dashed line). For 1300 gates or more, all fluctuation curves are visually indistinguishable from the Haar-8 benchmark.}
  \label{fig:plots_stdF_Q8}
\end{figure}

We now apply the same majorization-based characterization procedure to the 
Rigetti 8-qubit Agave QPU \cite{Reagor_2018}, which has a ``ring'' geometry with nearest-neighbor connections (see Fig.~\ref{fig:plots_Rigetti_topologies}(a)). 
Figure~\ref{fig:plots_stdF_Q8} shows the characteristic fluctuations computed for an ensemble of 5000 random quantum circuits and for increasing number of quantum gates. The dot-dashed curve is the limiting result for Cliff-8, which sets a limit for 8-qubit random Clifford circuits.
The dashed curve  is the result for Haar-8, which sets a lower limit for the fluctuations of 8-qubit random quantum circuits. Unlike what was observed for the 5-qubit IBM results above, for circuits composed of 100 and 300 gates, the fluctuations are significantly above those typical of  Cliff-8, indicating a lower level of complexity. As one would expect, generally, the larger 8-qubit Rigetti QPU requires a larger number of quantum gates to reach a level of complexity comparable to the Cliff-8 benchmark (500 gates for the Agave QPU). Though clearly distinguishable from each other, both curves are similar in shape and height. As the number of gates increases further to 700 and 900, however, we find a similar behavior to that of the IBM processors: as circuit complexity continues to grow, fluctuations decrease below the Clifford mark and approach the Haar-8 limit. Note that fluctuations for 1100 gates nearly coincide with Haar-8; for 1300 gates or more, the Lorenz fluctuaction curves become visually indistinguishable from Haar-8.

\begin{figure}[htb]  
  \centering
    \includegraphics[width=.5\textwidth]{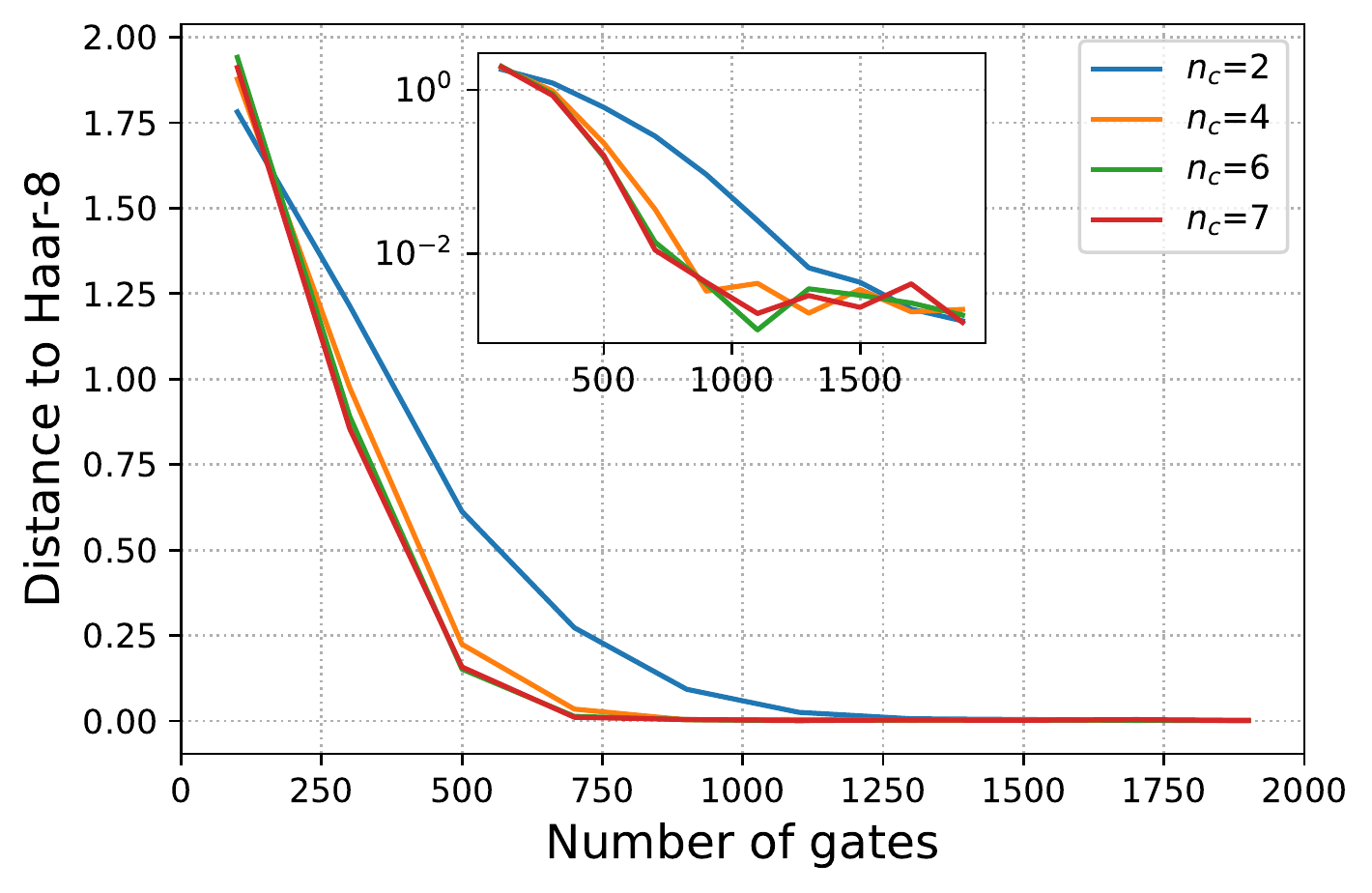}
  \caption{Distance to Haar-8, as defined in Eq.~(\ref{eq:DH}), as a function of the number of quantum gates for all 8-qubit QPU designs shown in Fig.~\ref{fig:plots_Rigetti_topologies}, with the Rigetti native gate set. The inset shows the same results in logarithmic scale. As expected, the distance to Haar-8  decreases more rapidly for  configurations  with  $n_c > 2$. Coincidence with Haar-8 is reached at 700 gates for the $n_c = 6$ and $n_c = 7$ configurations, 900 gates for the $n_c = 4$ geometry, and 1300 for the Agave ($n_c = 2$). Thus, due to the low connectivity of the Agave QPU, a significant larger number of quantum gates is necessary for reaching a high level of complexity.} 
  \label{fig:haar_distance_connectivity_Q8}
\end{figure}

The rather low nearest-neighbor qubit connectivity of the 8-qubit Rigetti Agave QPU ($n_c = 2$) motivates us to investigate the behavior of the cumulant fluctuations for  topologies with higher connectivity. Thus, using the Rigetti gate set $G_{\rm Rig} = \{\text{RX, RZ, CZ}\}$, we simulated three additional 8-qubit QPU configurations, with $n_c = 4, 6$ and 7 (all-to-all connectivity), which are illustrated in Figs.~\ref{fig:plots_Rigetti_topologies}(b)--(d), respectively. Results for the computed distances to Haar-8, as defined by Eq.~(\ref{eq:DH}), as a function of the number of quantum gates are shown in Fig.~\ref{fig:haar_distance_connectivity_Q8}. Notice, in particular, that the majorization-based indicator correctly captures the expected behavior of the distance to Haar-8 more rapidly as qubit connectivity increases. This, in turn, confirms that because of the low connectivity of the Agave QPU, a significantly larger number of quantum gates is necessary for reaching a high level of complexity. 
Fluctuations for the $n_c = 6$ and $n_c = 7$ configurations coincide with Haar-8 for 700 gates, 900 gates for the $n_c = 4$ geometry, and 1300 for the Agave ($n_c = 2$).

\section{Noisy quantum processors} 
\label{sec:noisy_quantum_processors}

We now turn to the question of how reliable is the majorization-based indicator when used for benchmarking a noisy quantum processor.

\subsection{Near-perfect QPUs} 
\label{sub:near_perfect_qpus}

 We start by considering the optimistic scenario of near-perfect QPUs. In this case, we assume that all gates can be applied perfectly. However, if a qubit is left idle, it experiences noise in the form of amplitude damping or dephasing. For simplicity, we consider each type of noise independently. Thus, following standard models~\cite{nielsen_chuang, preskill}, the Kraus operators for amplitude damping are 
\begin{align}
D_0 & = \ket{0}\bra{0} + \sqrt{1-p}\ket{1}\bra{1}, \\
D_1 & = \sqrt{p}\ket{0}\bra{1}, 
\end{align}
where $p = 1-\exp(-t/T_1)$, and for pure dephasing they are given by
\begin{align}
\tilde D_0 &=  \sqrt{1-\tilde p} \, I, \\
\tilde D_1 &=  \sqrt{\tilde p} \, Z, 
\end{align}
with $\tilde p= 1/2 (1-\sqrt{1-\exp(-2t/T_2)})$. 
Note that the noise strengths are solely set by the characteristic times $T_1$ and $T_2$, respectively. 

\begin{figure}[th]  
  \centering
    \includegraphics[width=.5\textwidth]{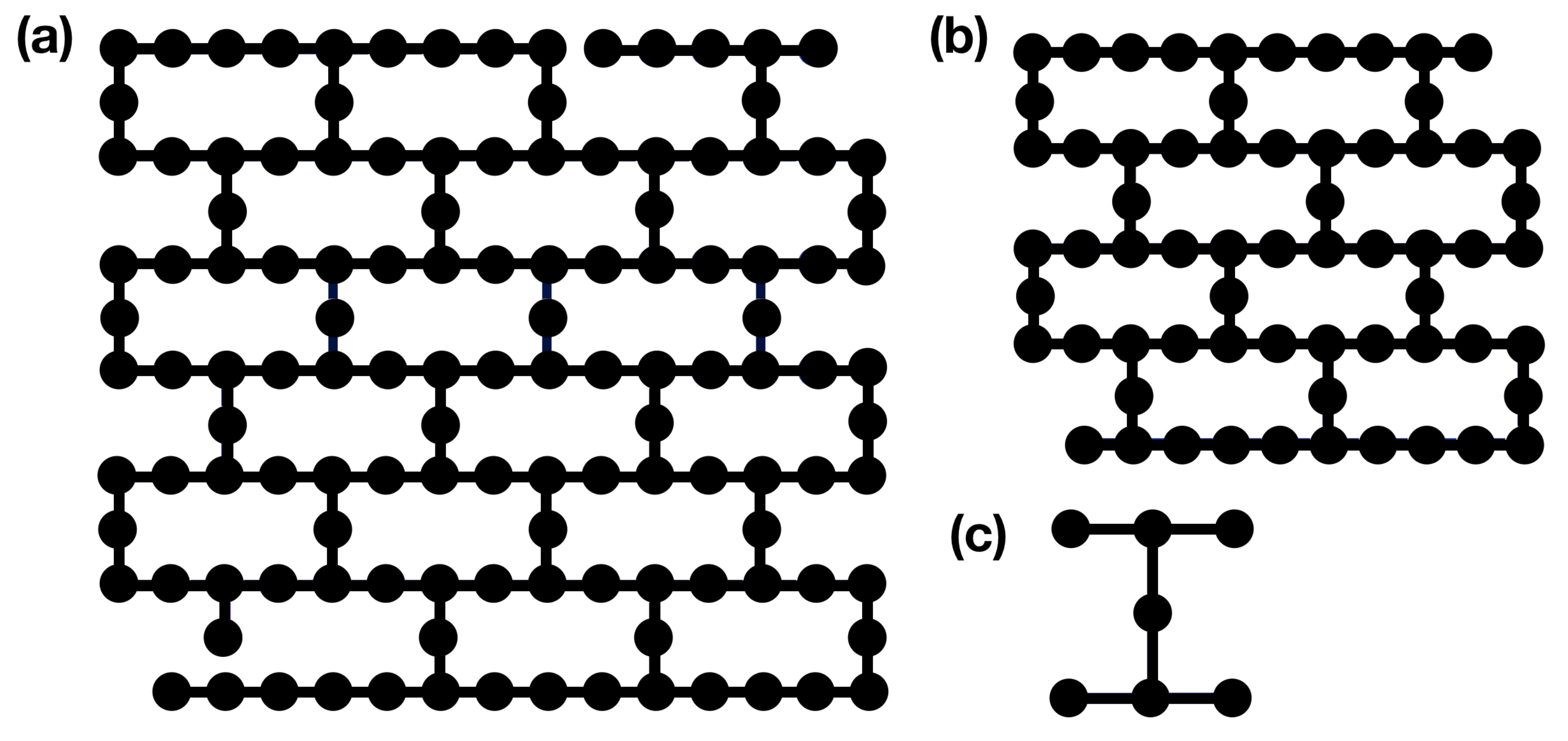}
  \caption{Examples of currently available IBM QPU geometries~\cite{ibm_url}: (a) IBM Washington with 127 qubits, (b) IBM Ithaca with 65 qubits, and (c) IBM Perth with 7 qubits. Note that the geometry of the Perth QPU also corresponds to 7-qubit segments (sub-sections) of the design of the larger QPUs.}
  \label{fig:ibm_qpu_designs}
\end{figure}

According to the calibration data available at IBM's quantum computing resources website~\cite{ibm_url}, $T_1$ and $T_2$ can vary significantly among different processor families, QPU implementations, and individual qubits. For instance, in the Eagle family (currently the most advanced one), the IBM Washington QPU has 127 qubits with average noise times of the order of 100~$\mu$s. Depending on specific qubits, $T_1$ ranges approximately from 25 to 175~$\mu$s and $T_2$ from 3 to 270~$\mu$s. For the Hummingbird family, the IBM Ithaca QPU (Hummingbird, 65 qubits), for instance, has reported average noise times of the order of 200~$\mu$s, with $T_1$ ranging approximately from 70 to 315~$\mu$s and $T_2$ from 22 to 500~$\mu$s. Falcon processors have also similar noise time scales, despite their smaller sizes of up to 27 qubits. Given such a wide range of QPU designs and noise strengths for currently available IBM quantum computers, we consider below the case of 7 qubits interconnected as an ``H'', as illustraded in Fig.~\ref{fig:ibm_qpu_designs}. This qubit arrangement is particularly appealing for our discussion, as it can be found in all those IBM quantum processor families, either as a 7-qubit standalone QPU (for instance, the IBM Perth, Falcon), or as a 7-qubit segment (i.e., a sub-section) of a larger QPU (see Fig.~\ref{fig:ibm_qpu_designs}). Following realistic gate-time data available through Qiskit~\cite{qiskit}, we consider the following gate times for our simulations: 36~ns for $\sqrt{X}$, 400~ns for CNOT, while RZ rotations are assumed to be instantaneous. For simplicity, we set the same gate and noise times across all qubits.

\begin{figure}[h]   
  \centering	
    \includegraphics[width=.5\textwidth]{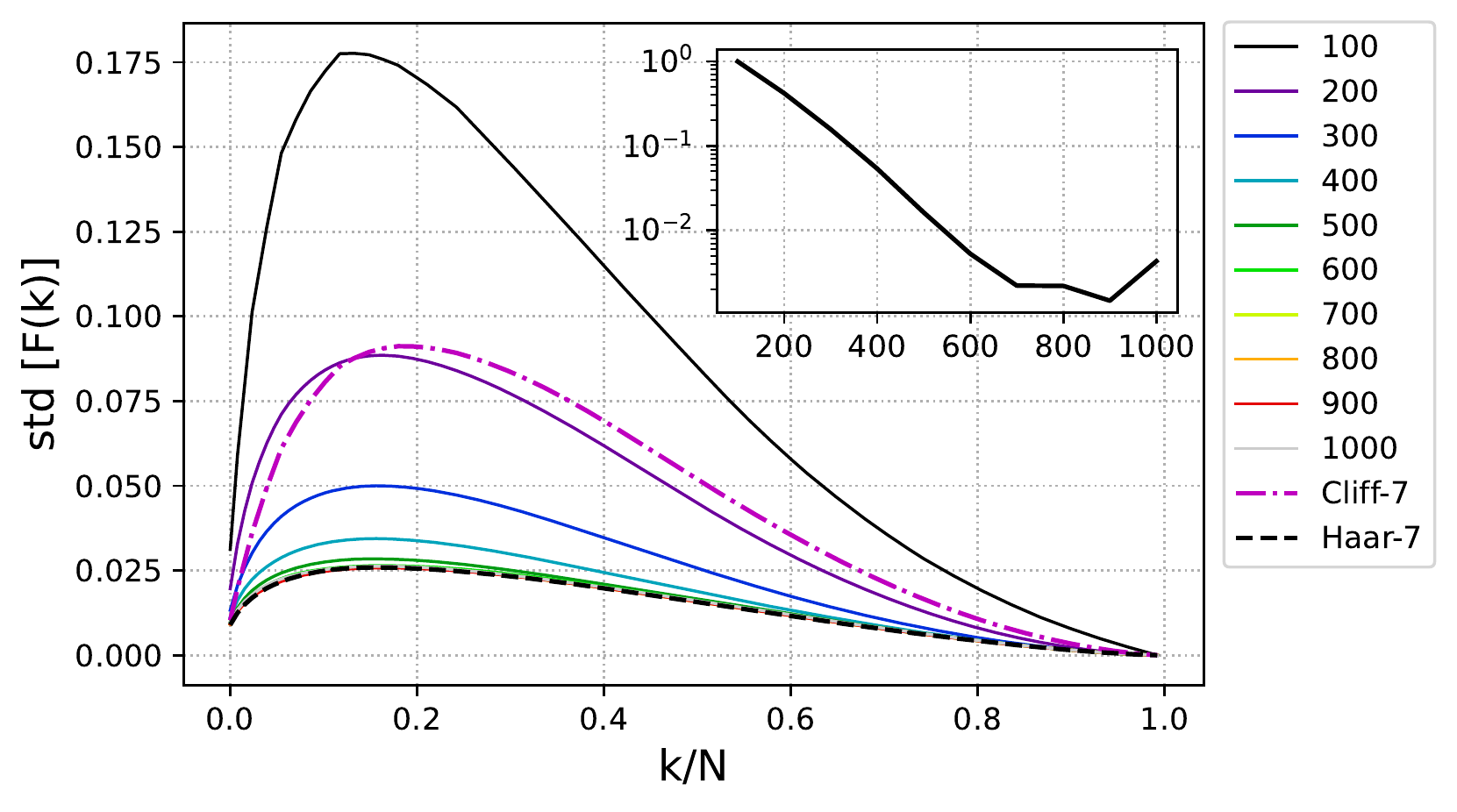}
  \caption{Characterization of a \emph{noiseless}, 7-qubit, “H”-connected IBM QPU for 5 000 random circuits and increasing number of gates. The dot-dashed curve is the result for Cliff-7, i.e., the limiting curve for 7-qubit random Clifford circuits. The dashed line is the reference curve for Haar-7, which represents the complexity of Haar-random 7-qubit pure states. For 100 gates, fluctuations are significantly above the Cliff-7, indicating a lower complexity. For 200 gates, fluctuations are comparable to Cliff-7 and for 300 gates or more, all fluctuations curves approach the Haar-7 line and become visually indistinguishable for 600 gates or more. This is can also be seen from the inset plot, which shows the distance to Haar-7, defined in Eq.~\ref{eq:DH}, as a function of the number of quantum gates applied.}
  \label{fig:plots_ibm_H7_pure}
\end{figure}

First, we perform the characterization of the \emph{noiseless} IBM Perth QPU (or a 7-qubit, ``H''-connected  segment), which is shown in Fig.~\ref{fig:plots_ibm_H7_pure} for 5000 random circuits. For random circuits with 100 gates, fluctuations are significantly larger than Cliff-7 reference line. For 200 gates, we see a small region of near-coincidence with Cliff-7 ($0 < k/N \lesssim 0.1$). For $k/N \gtrsim 0.1$, fluctuations are below Cliff-7. For 300, 400 and 500 gates, however, there is a visible separation from Cliff-7 and fluctuations are much closer to the Haar-7 benchmark line. For 600 gates or more, the device's characteristic curves are visually indistinguishable from Haar-7. This is more accurately seen from the inset plot, which shows the distance~(\ref{eq:DH}) between each fluctuation curve and the Haar-7 line.

\begin{figure}[htb]  
  \centering
    \includegraphics[width=.5\textwidth]{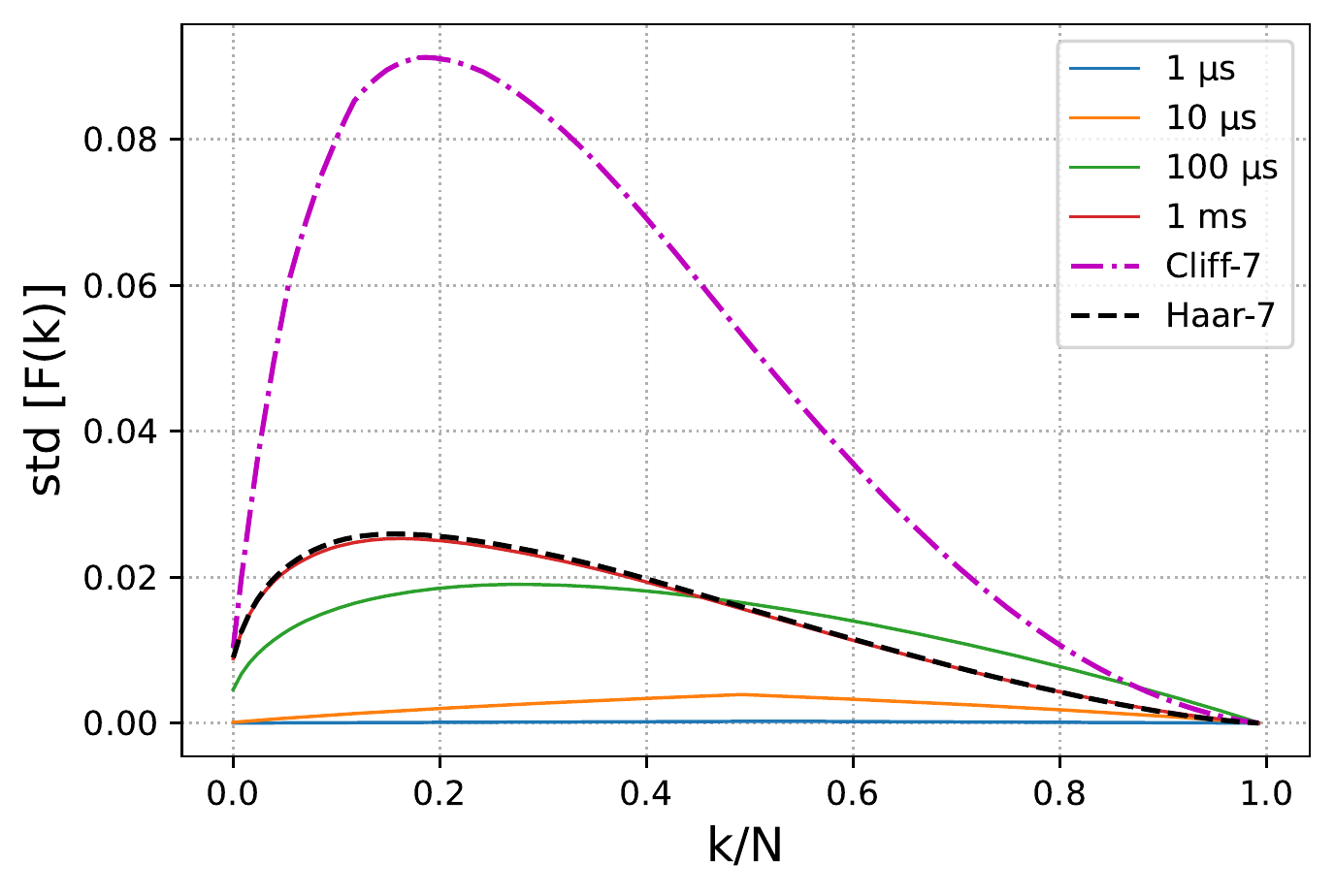}
  \caption{Fluctuations for 5000 random circuits running on a 7-qubit, “H”-connected IBM QPU with \emph{dephasing} and $T_2 = 1, 10, 100~\mu$s, and 1~ms (solid lines from bottom to top, respectively). The reference curves are the same as shown before for the noiseless cases. 
  The dot-dashed curve is the limiting curve for 7-qubit random Clifford circuits, whereas the dashed line represents the complexity of Haar-random 7-qubit pure states. 
  For $T_2 = 1~\mu$s, fluctuations are vanishingly small. 
  In this case, noise is too strong and the output states approach the maximally mixed state (see average purity in Fig.~\ref{fig:Rigetti_fid_purities}). For $T_2 = 10~\mu$s, fluctuations are still close to zero, indicating a very strong noise regime. For $T_2 = 100 \mu$s, which is the relevant noise time scale for current IBM processors, fluctuations are somewhat comparable to Haar-7. Although, obviously distinguishable from each other, both curves have a similar shape and height. This suggests that quantum complexity may have not been fully washed out by noise (as shown in Fig.~\ref{fig:plots_ibm_H7_fidelity}, the corresponding average purity is $\approx 0.25$ and the average fidelity is $\approx 0.5$). For $T_2 = 1$~ms, the noise strength is low enough to produce no apparent deviation from the Haar-7 benchmark line.}
  \label{fig:plots_ibm_H7_deph}
\end{figure}

\begin{figure}[tbp]  
  \centering
    \includegraphics[width=.5\textwidth]{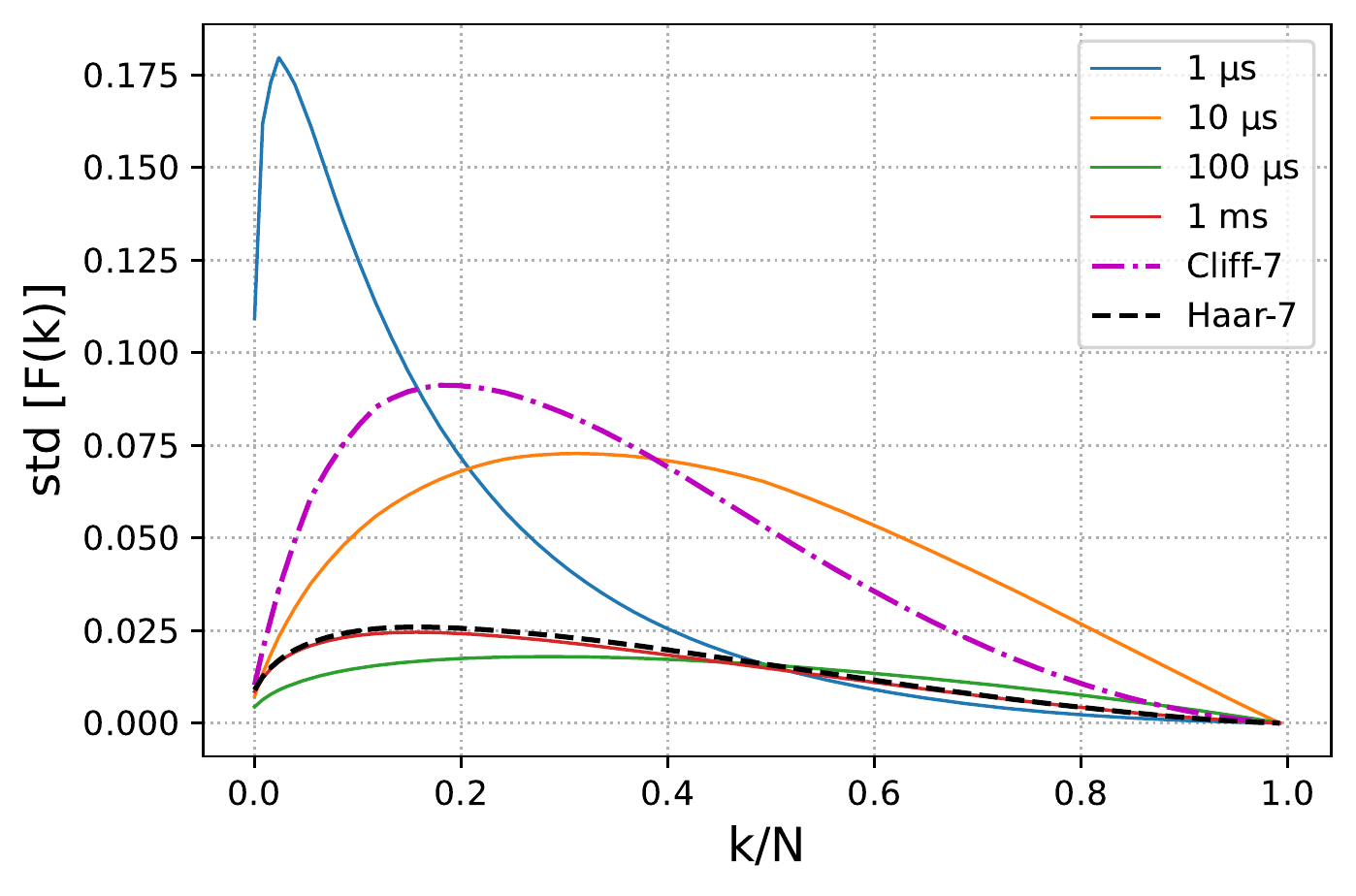}
  \caption{Fluctuations for 5000 random circuits running on a 7-qubit, “H”-connected IBM QPU with \emph{amplitude damping} and $T_1 = 1, 10, 100~\mu$s, and 1~ms (solid lines from bottom to top, respectively). The reference curves Cliff-7 and Haar-7 are the same as shown before for the noiseless cases. 
  Note that fluctuations here are qualitatively different than the case of pure dephasing. 
  For $T_1 = 1~\mu$s, fluctuations are significantly higher than Cliff-7, with a narrow peak close to $k/N \approx 0$. This is a strong noise regime, 
  with all the variance being due to only a small number of components. This is in agreement with the corresponding results in Fig.~\ref{fig:plots_ibm_H7_fidelity}: zero average fidelity and average purity $\approx 0.2$. For $T_1 = 10~\mu$s, fluctuations become spread out over the entire $k/N$-domain and are comparable with Cliff-7, having similar shape and height. For $T_1 = 100 \mu$s, which is the relevant noise time scale for current IBM processors, fluctuations indicate a complexity somewhat comparable to Haar-7, as both curves resemble in shape and height. For $T_1 = 1$~ms, the noise strength is low enough to produce no apparent deviation from the Haar-7 benchmark line.}
  \label{fig:plots_ibm_H7_damp}
\end{figure}

Now we fix the number of gates to 600 and include only one type of noise at a time. 
Fluctuations for 5000 random circuits running on a noisy QPU with dephasing are shown in Fig.~\ref{fig:plots_ibm_H7_deph} and with amplitude damping in Fig.~\ref{fig:plots_ibm_H7_damp}. 
For both noise models, we vary the noise time scales from 1~$\mu$s to 1~ms. 

\begin{figure}[htbp]  
  \centering
    \includegraphics[width=.5\textwidth]{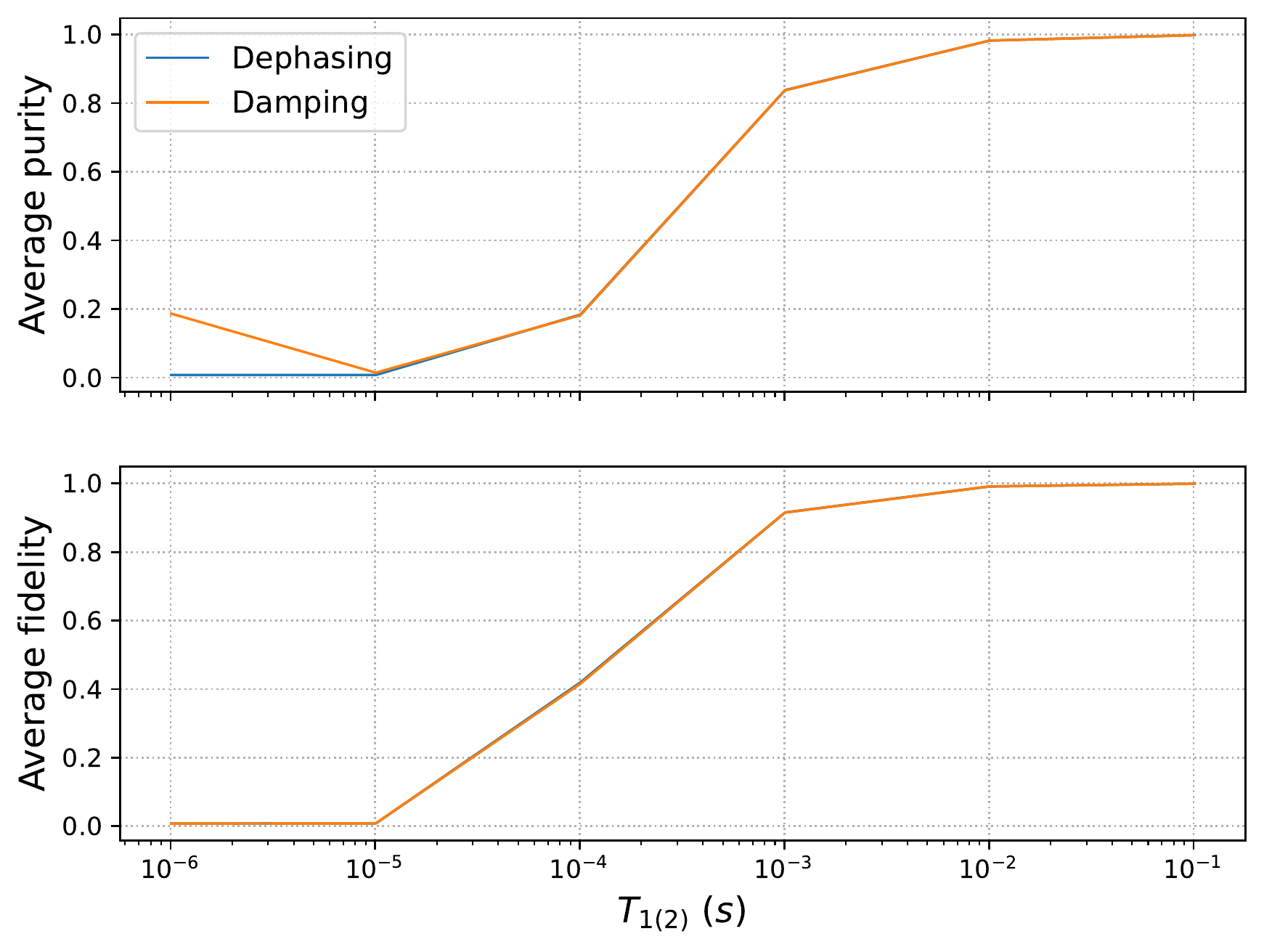}
  \caption{Average purity/fidelity over random circuits running on a 7-qubit, 
  “H”-connected IBM QPU. (top) Average purity of the circuits’ output states (pre-measurement) as a function of noise times $T_1$ and $T_2$. (bottom) Average fidelity between noisy and noiseless output states (for the same circuit) as a function of $T_1$ and $T_2$. Averages were done over for an ensemble of 5000 circuits with 600 gates. These plots show three main noise regimes. For $T_{1 (2)} < 10~\mu$s, we see that noise is very strong, leading to vanishing low values of fidelity and purity. At the other extreme, for $T_{1 (2)} \gtrsim 10$~ms, noise is very low, allowing for computations with near-one fidelity and purity. For $10~\mu$s $\lesssim T_{1 (2)} < 1~\mu$s, we see an intermediate noise regime, in which there is a rapid increase from low to high values of purity and fidelity. As shown in Figs.~\ref{fig:plots_ibm_H7_deph} and \ref{fig:plots_ibm_H7_damp}, visual coincidence of the fluctuations of the Lorentz curves with the Haar-7 benchmark line is only achieved for $T_{1 (2)} \gtrsim 1~\mu$s, i.e., a regime of high fidelity and purity.}
  \label{fig:plots_ibm_H7_fidelity}
\end{figure}

To have a clearer understanding of the behavior of the indicator in those different noise regimes, 
we show in Fig.~\ref{fig:plots_ibm_H7_fidelity} the average purity of the circuits' output states (pre-measurement) and also the average fidelity between the noisy and the noiseless output states (for the same random circuit). 
For $T_{1 (2)} = 1~\mu$s, we see a strong noise regime for both noise models, 
despite the significant differences between those characteristic curves. 
For the case of dephasing, fluctuations vanish and so do the average purity and the average fidelity. 
These results evidence that all output states have decohered into a maximally mixed state. Note that, in the presence of noise, Haar-7 (more generally Haar-$n$) is no longer a limiting curve. On the other hand, fluctuations for the case of amplitude damping have a narrow peak (width at half maximum $\approx$ 0.15) that is nearly two times higher than the Cliff-7 line. Although the average purity being approximately 0.2 indicates that there is some residual quantum coherence left in the system, the average fidelity is zero. For $T_{1 (2)} = 10~\mu$s, fluctuations in Fig.~\ref{fig:plots_ibm_H7_deph} are still close to zero and in Fig.~\ref{fig:plots_ibm_H7_damp} are nearly as large as Cliff-7. 
Therefore, for both models, noise is still too large to allow achieving a higher level of complexity.
This claim is further supported by the fact that both average purity and average fidelity are zero. Interestingly, we see a significant qualitative change for $T_{1 (2)} = 100~\mu$s, which is precisely the relevant noise time scale for current IBM processors. The characteristic curves for both noise types are now comparable to Haar-7. 
Fluctuations are maxed out almost at the same value and both curves do resemble the Haar-7 line in shape. However, they are obviously distinguishable from each other. Therefore, this represents an intermediate noise regime, in which quantum complexity has not been fully degraded. Indeed, the average purity is approximately 0.25 and the average fidelity is 0.5. Only for $T_{1 (2)}$ = 1~ms we reach a low noise regime, where there is no apparent deviation from the Haar-7 benchmark line. This coincidence for both noise types corresponds to an average purity of approximately 0.86 and a remarkable average fidelity of 0.93.

The results above show that this majorization-based characterization procedure yields meaningful results also in the presence of noise, as long as the circuits' output states have high purity on average. To verify this claim further, we now apply the same benchmark procedure to the case of noisy Rigetti QPUs. According to Rigetti's website \cite{rigetti_url}, the values of $T_{1}$ and $T_{2}$ for their processors also vary among different families and QPU implementations. The M-3, M-2 and M-1 processors in the Aspen family, equipped with 80 qubits, have median values of $T_{1}$ ranging from 22~$\mu s$ to 31~$\mu s$, and median values of $T_{2}$ ranging from 18~$\mu s$ to 24~$\mu s$. For the Agave 8-qubit device, the average value of $T_{1}$ is 13.38~$\mu s$, and the average value for $T_{2}$ is 15.05 $\mu s$. In a more detailed paper showing an experiment using the Agave 8-qubit device \cite{Reagor_2018}, it was shown that $T_{1}$ varies for each qubit, ranging from 34.1~$\mu s$ to 5.6~$\mu s$. Meanwhile, $T_{2}^*$ ranges from 18.7~$\mu s$ to 4.3~$\mu s$.

First, we analyze the cumulant fluctuations of a noisy Rigetti 8-qubit Agave QPU --- whose noiseless case was analyzed in Sec.~\ref{sec:noiseless_quantum_processors}. According to \cite{rigetti_url}, the average time for one-qubit gates for this device is 50 ns, while the average time for two-qubit gates is 160 ns. In our simulations, we used 50 ns as the duration time of one-qubit gates and approximated to 150 ns the duration time of two-qubit gates.

\begin{figure}[htbp] 
  \centering
    \includegraphics[width=.5\textwidth]{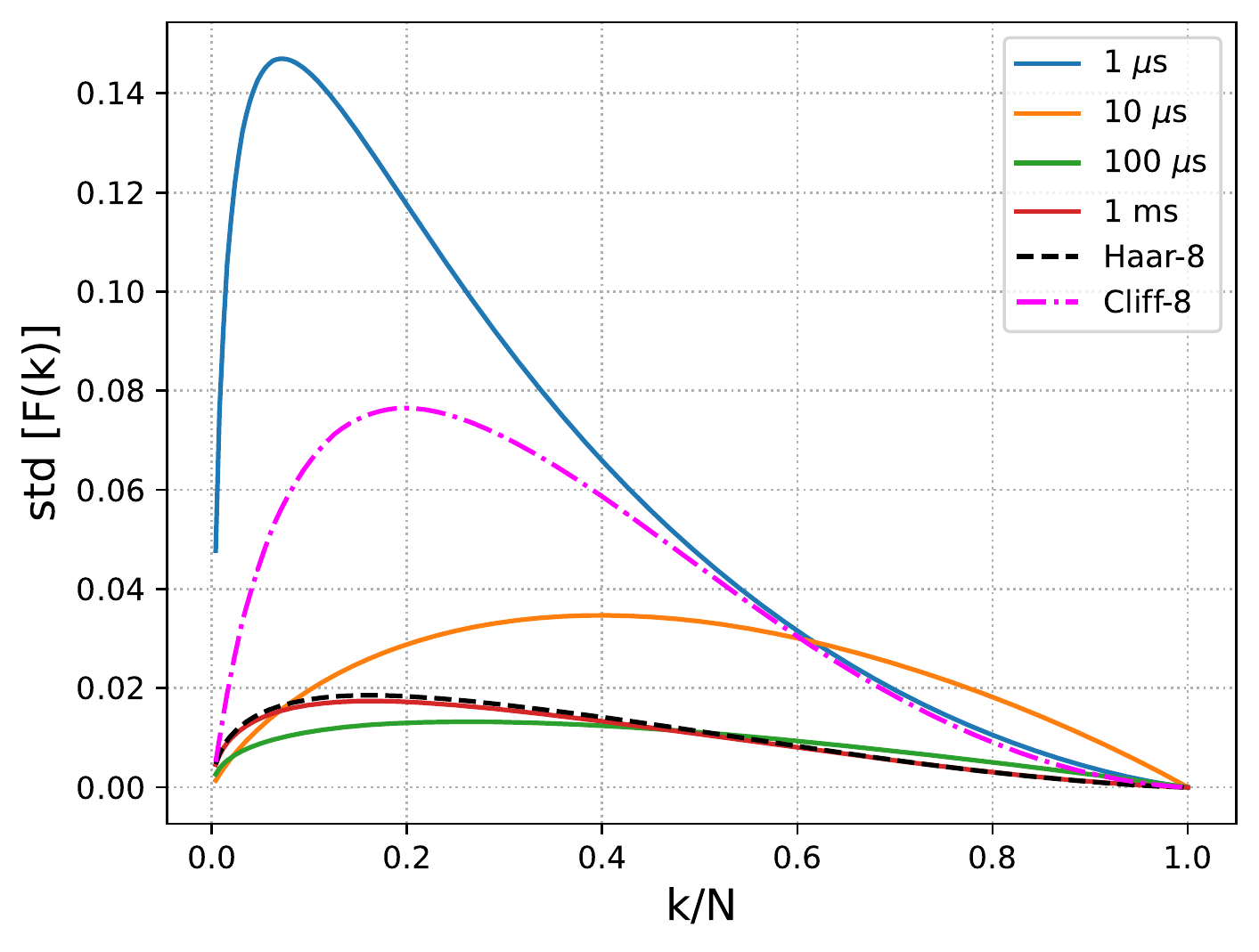}
  \caption{Majorization-based characterization of a noisy Rigetti Q8 Agave device with amplitude damping and $T_1 = 1, 10, 100, \text{and }  1000~\mu$s (solid lines from bottom to top, respectively). Here we plot the fluctuations of the Lorenz curves for 5000 random circuits with a fixed number of 1500 gates. The reference curves Cliff-8 and Haar-8 are the same as used before for the noiseless cases.  
  For $T_1 = 1~\mu s$, fluctuations are peaked significantly above Cliff-8. This is characteristic of a strong noise regime, which is in agreement with the respective results shown in Fig.~\ref{fig:Rigetti_fid_purities}: zero average circuit fidelity and near-zero average purity. For $T_1 = 10~\mu$s, fluctuations become spread out over all values of $k/N$, but the change in shape (relative to Haar-8) is due to a significant amount of noise in the computation. This is also supported by the fact that both average purity and average fidelity are zero. Indeed, note that, once noise is further reduced, by setting $T_1 = 100~\mu$s, the characteristic curves become comparable to Haar-8 in shape. This represents an intermediate noise regime, with average purity 0.28 and average fidelity 0.52. A low noise regime emerges for $T_1$ = 1~ms, with no apparent deviation from Haar-8 (average purity $\approx$ 0.87 and average fidelity $\approx$ 0.93).}
  \label{fig:T1_error_fluctuations}
\end{figure}

\begin{figure}[htbp]  
  \centering
    \includegraphics[width=.5\textwidth]{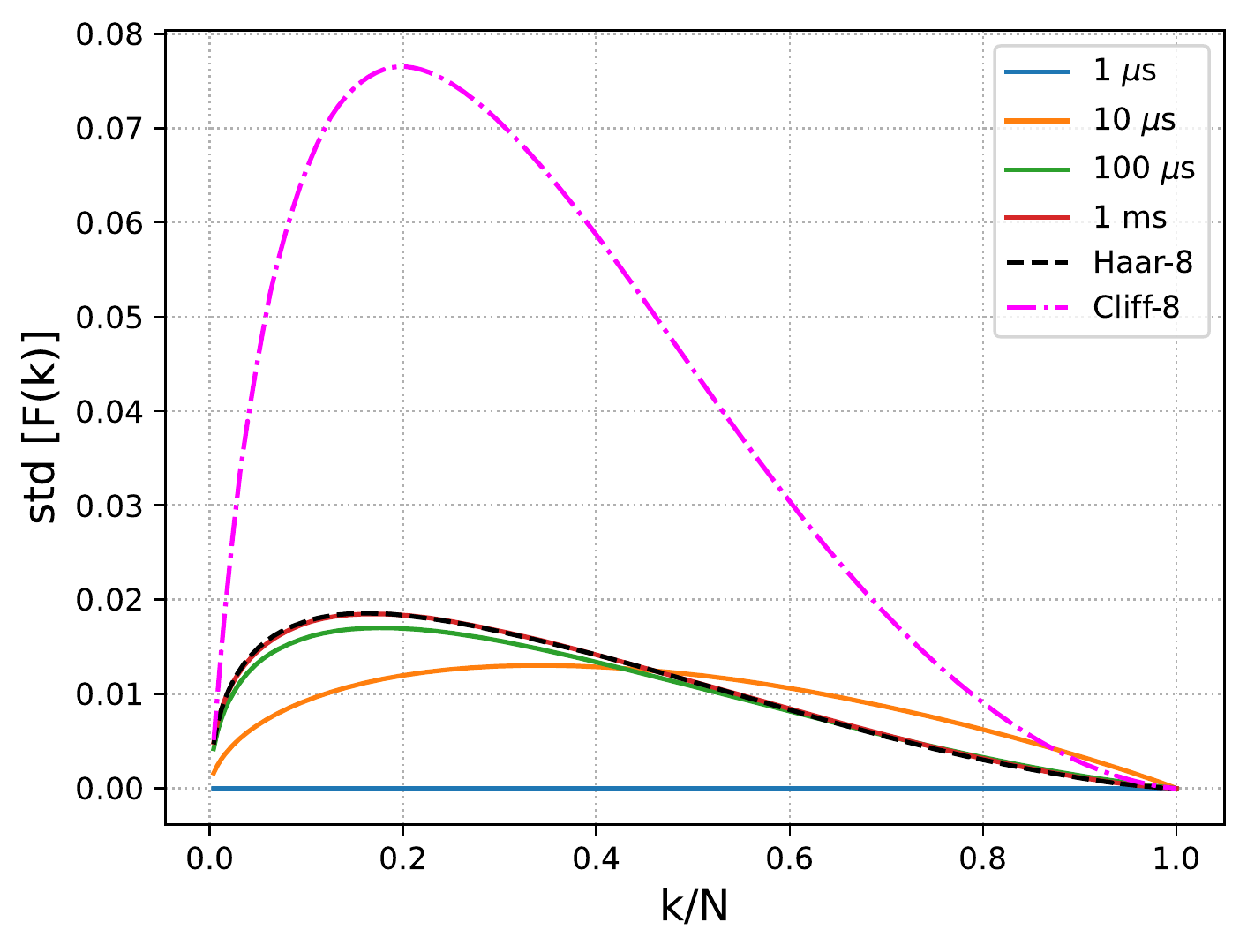}
  \caption{Majorization-based characterization of a noisy Rigetti Q8 Agave device with pure dephasing and $T_2 = 1, 10, 100, \text{and } 1000~\mu$s (solid lines from bottom to top, respectively). 
  Here we plot the fluctuations of the Lorenz curves for 5000 random circuits with a fixed number of 1500 gates. The reference curves Cliff-8 and Haar-8 are the same as used for the noiseless cases. For $T_2 = 1~\mu$s, due to strong noise, fluctuations are vanishingly small, and so does the average circuit fidelity and the average purity (see Fig.~\ref{fig:Rigetti_fid_purities}). For $T_2 = 10~\mu$s, fluctuations are no longer zero and become spread out over all values of $k/N$. However, the difference in shape relative to Haar-8 reveals the presence of a significant amount of noise (note that both average purity and average fidelity remain zero). For $T_2 = 100~\mu$s, the Lorentz curve become comparable to Haar-8 in shape, indicating an intermediate noise regime (average purity $\approx$ 0.28 and average circuit fidelity $\approx$ 0.52). A low noise regime is only achieved when $T_2 = 1$~ms. In this case, there is visual coincidence with Haar-8 (average purity $\approx$ 0.87  and average circuit  fidelity $\approx$ 0.93).}
  \label{fig:T2_error_fluctuations}
\end{figure}

\begin{figure}[htbp]  
  \centering
    \includegraphics[width=.5\textwidth]{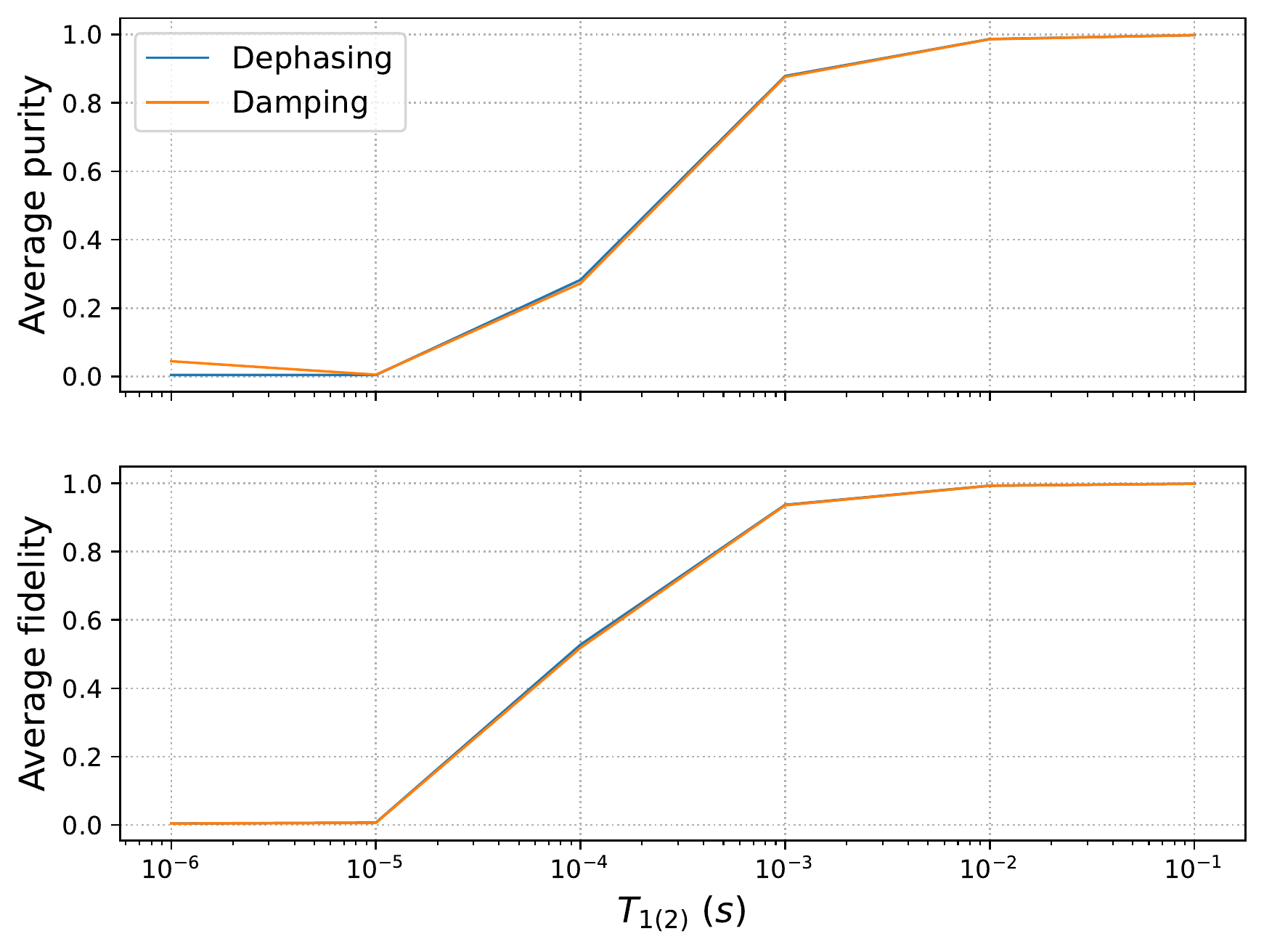}
  \caption{(top) Average purity of the circuits’ output states (pre-measurement) as a function of noise times $T_1$ and $T_2$ for the Rigetti Q8 Agave processor. (bottom) Average fidelity between noisy and noiseless output states (for the same circuit) as a function of $T_1$ and $T_2$. Averages done over an ensemble of 5000 circuits with 1500 gates. For $T_{1 (2)} < 10~\mu$s, noise is very strong, leading to vanishing low values of purity and  circuit fidelity. For $10~\mu$s $\lesssim T_{1 (2)} \lesssim 1~\mu$s, we see an intermediate noise regime, corresponding to a rapid increase from low to high values of purity and fidelity. For $T_{1 (2)} \gtrsim 10$~ms, noise is very low, allowing for computations with near-one fidelity and purity.}
  \label{fig:Rigetti_fid_purities}
\end{figure}

As before, we assume that all quantum gates can be applied perfectly, but idle qubits are affected by noise, either in the form of amplitude damping or pure dephasing. For both noise models, we varied the noise time scales from 1~$\mu$s to 1~ms and simulated ensembles of 5000 random circuits with a fixed number of 1500 gates. Figure~\ref{fig:T1_error_fluctuations} shows the results  when the only type of noise present is amplitude damping, while Fig.~\ref{fig:T2_error_fluctuations} shows the results for pure dephasing. Additionally,  Fig.~\ref{fig:Rigetti_fid_purities} shows the average purity of the circuits' output states and the average circuit fidelity between the noisy and the noiseless output states. 

Despite having completely different architectures, the results we find for the noisy Rigetti Agave are remarkably similar to the ones above for the IBM Perth, \sout{e}specially the ones for amplitude damping. First, notice that for $T_{1 (2)} = 1~\mu s$, we again see features of a strong noise regime for both noise models. For the case of dephasing, fluctuations are vanishingly small, whereas for the case of amplitude damping, the cumulant fluctuations are above  Cliff-8, having a narrow peak near $k/N \approx 0$ (i.e., all the variance is mostly due to a few components). Both cases have zero average circuit fidelity and average purity (near-zero for amplitude damping), as shown in Fig.~\ref{fig:Rigetti_fid_purities}. For $T_{1 (2)} = 10~\mu$s, fluctuations become spread out over all values of $k/N$. However, the difference in shape relative to Haar-8 reveals the presence of a significant amount of noise. This is also supported by the fact that both average purity and average circuit fidelity are zero. Indeed, note that, by further reducing the amount of noise, by setting $T_{1 (2)} = 100~\mu$s, the characteristic curves become comparable to Haar-8 in shape, though being visually distinguishable from it. As discussed before, this represents an intermediate noise regime, with average purity $\approx$ 0.28 and average fidelity 0.52. Lastly, a low noise regime emerges for $T_{1 (2)}$ = 1~ms, as there is no apparent deviation from Haar-8 (average purity $\approx$ 0.87  and average fidelity $\approx$ 0.93).

\subsection{Faulty QPUs} 
\label{sub:faulty_qpus}

In a more realistic scenario, quantum gates may suffer from imperfections, in addition to idle noise. In this section, we investigate the case of faulty QPUs, by treating gate imperfections as random Pauli errors. In this case, we describe the effect of single-qubit and two-qubit errors on the processor's quantum state by means of depolarizing channels~\cite{preskill}. The Kraus operators for one qubit are given by
\begin{eqnarray*}
K_0 & =\sqrt{1-\varepsilon_1} I, \\
K_1 & =\sqrt{\varepsilon_1 / 3} X, \\
K_2 & =\sqrt{\varepsilon_1 / 3} Y, \\
K_3 & =\sqrt{\varepsilon_1 / 3} Z.
\end{eqnarray*}
For two qubits, we take the Kraus operators to be tensor products of the Pauli operators $\{I, X, Y, Z\}$, with the coefficient $\sqrt{1-\varepsilon_2}$ for the identity and $\sqrt{\varepsilon_2/15}$ for each remaining operator. Note that, in this way, the error rates $\varepsilon_1$ and $\varepsilon_2$ control the strength of each noise type independently. 
In order to analyze how robust the majorization-based indicator is against these types of error, we do not add idle noise to the simulations presented below. 
Since the results of the previous section were qualitatively similar for the IBM and Rigetti processors, for brevity, we focus here on the case of a faulty, 7-qubit, “H”-connected IBM QPU only.

\begin{figure*}[htbp]  
  \centering
    \includegraphics[width=\textwidth]{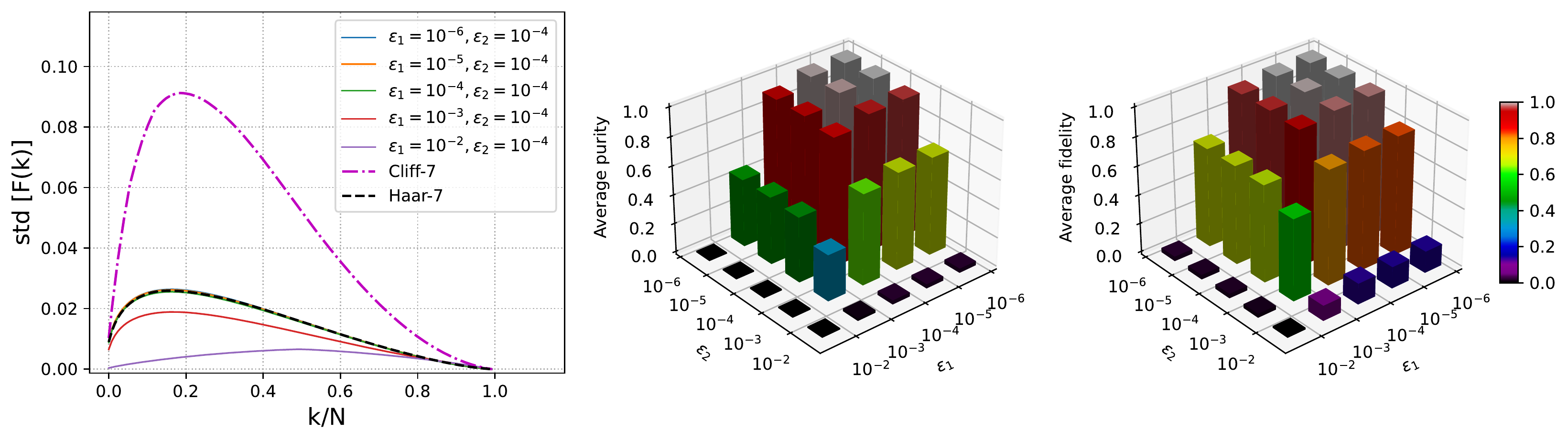}
  \caption{(left) Majorization-based characterization of a faulty, 7-qubit, “H”-connected IBM QPU for an ensemble of 20000 random circuits with 600 gates. Here we set $\varepsilon_2=10^{-4}$ and vary $\varepsilon_1$. The reference curves Cliff-7 and Haar-7 are the same as used before. For $\varepsilon_1 \leq  10^{-4}$, all fluctuation curves coincide with Haar-7. For larger error rates, i.e., $\varepsilon_1 = 10^{-3}$ and $10^{-2}$, fluctuations decrease, falling below the Haar-7 line. (center) Average purity of the circuits’ output states (pre-measurement) as a function of the error parameters $\varepsilon_1$ and $\varepsilon_2$. (right) Average fidelity between noisy and noiseless output states as a function of $\varepsilon_1$ and $\varepsilon_2$. The purpose of the colorbar here is to highlight the low noise regime, in which both (average) purity and fidelity are high. Precise values are shown in Table~\ref{tab:depolarazing}. A low noise regime corresponds to $\varepsilon_{1 (2)} \lesssim 10^{-4}$.  For larger values of either $\varepsilon_1$ or $\varepsilon_2$, both purity and fidelity rapidly decreases.}
  \label{fig:plots_ibm_H7_depol}
\end{figure*}

Thus, following the same protocol as before, we fix the number of gates to 600 and include both one- and two-qubit errors, with $\varepsilon_{1 (2)}\in \{10^{-2}, 10^{-3}, 10^{-4}, 10^{-5}, 10^{-6}\}$. Figure~\ref{fig:plots_ibm_H7_depol} shows the results for the simulation of an ensemble of 20000 random circuits. 
Here it was necessary to increase the size of the ensemble of random circuits to deal with small numerical errors. On the left panel, we show the fluctuations of the Lorenz curves for increasing error rates. The reference curves Cliff-7 and Haar-7 are the same as used before. To avoid confusion, instead of showing all simulated Lorenz curves, we focus on the more instructive case of fixed $\varepsilon_2 = 10^{-4}$ and variable $\varepsilon_1$. As we increase $\varepsilon_1$, we can clearly observe a transition from a low to a strong noise regime. 
For $\varepsilon_1 \leq  10^{-4}$, all fluctuation curves coincide with Haar-7. For larger error rates, i.e., $\varepsilon_1 = 10^{-3}$ and $10^{-2}$, fluctuations decrease, falling below the Haar-7 line. Note that this is a similar effect to the case of strong dephasing shown in Fig.~\ref{fig:plots_ibm_H7_deph}. As the error rates increase, so does the coefficient of the identity component of the output state of the depolarizing map, which, in turn, results in a decrease of the cumulant fluctuations and of the effective circuit complexity.

We again computed the average purity of the circuits’ output (pre-measurement) states and the average circuit fidelity between the noisy and the noiseless output states, which we show on the middle and right panels, respectively, for all $\varepsilon_1$ and $\varepsilon_2$ considered. Here, we have included a color-bar only to help visualization of the low noise regime, corresponding to high values of average purity and average circuit fidelity. For clarity, we also show the precise values in Table~\ref{tab:depolarazing}. From those, it becomes clear that the low noise regime corresponds to $\varepsilon_{1 (2)} \lesssim 10^{-4}$.  For larger values of either $\varepsilon_1$ or $\varepsilon_2$, both purity and circuit fidelity rapidly decreases. Therefore, despite the qualitative nature of the majorization-based characterization, a near-coincidence of the fluctuations with the Haar-7 line, corresponds to, on average, a near-one purity and fidelity.

\begin{table}[htbp]
    \centering
    \setlength\arrayrulewidth{1pt}
    \begin{tabular}{|c|c!{\vrule width 0.1pt}c|c!{\vrule width 0.1pt}c|c!{\vrule width 0.1pt}c|c!{\vrule width 0.1pt}c|c!{\vrule width 0.1pt}c|}
    \hline
    \diagbox{$\varepsilon_1$}{$\varepsilon_2$} &
      \multicolumn{2}{c|}{$10^{-6}$} &
      \multicolumn{2}{c|}{$10^{-5}$} &
      \multicolumn{2}{c|}{$10^{-4}$} &
      \multicolumn{2}{c|}{$10^{-3}$} &
      \multicolumn{2}{c|}{$10^{-2}$} \\
      \hline
      \rule{0pt}{.4cm}
         $10^{-6}$ & 99.9&99.9 & 99.5&99.8 & 96.1&98.0 & 67.9&82.4 & 2.9&14.9 \\\hline
      \rule{0pt}{.4cm}
         $10^{-5}$ & 99.2&99.6 & 98.9&99.4 & 95.5&97.7 & 67.4&82.1 & 2.9&14.9 \\\hline
      \rule{0pt}{.4cm}
         $10^{-4}$ & 92.6&96.2 & 92.3&96.1 & 89.1&94.4 & 63.0&79.3 & 2.7&14.4 \\\hline
      \rule{0pt}{.4cm}
         $10^{-3}$ & 46.9&68.3 & 46.7&68.2 & 45.1&67.0 & 32.0&56.4 & 1.8&10.5 \\\hline         
      \rule{0pt}{.4cm}
         $10^{-2}$ & 0.9&3.1 & 0.9&3.1 & 0.9&3.0 & 0.8&2.7 & 0.8&1.1 \\\hline         
\end{tabular}
    \caption{Average purity of the circuits’ output states and average fidelity between noisy and noiseless output states (values shown in this order and as percentages) for various error rates $\varepsilon_1$ and $\varepsilon_2$. Averages were done over for an ensemble of 20000 circuits with a fixed number of 600 gates.}
    \label{tab:depolarazing}
\end{table}



\section{Discussion} 
\label{sec:discussion}

In this article we have investigated the use of the majorization-based indicator introduced in Ref.~\cite{vallejos01} as a way to potentially benchmark the complexity within reach of currently available quantum processors, even in the presence of noise. Essentially, it accounts for computing the fluctuations of the Lorenz curves of the output probabilities (in the computational basis) of random quantum circuits, generated from the device's native gate set. To evaluate its performance, we applied this simple architecture-independent protocol to several available QPU architectures by IBM and Rigetti. We numerically simulated their operation under various conditions and characterized their complexity for various native gate sets, qubit connectivities, number of gates, and noise types. Key levels of complexity were identified by direct comparison to reference curves obtained for randomized Clifford circuits and Haar-random pure states. In this way, we were able to pin down, for each specific architecture, the number of native quantum gates which are necessary, on average, for achieving those levels of complexity. In addition, for noisy processors, our analysis showed that the majorization-based benchmark still holds for various types of noise as long as the circuits' output states have, on average, high purity ($\gtrsim 0.9$). This allowed us to identify a low noise parameter regime for each device, in which, the indicator showed no significant differences from the noiseless case. 

Note, however, that this approach may be further refined in order to accommodate more specific hardware constraints. For instance, one could include additional reference curves, whose complexities are known to be higher than Clifford (see Ref.~\cite{vallejos01} for several examples). Another possibility for dealing with noisy QPUs would be to have a noise model added to the reference lines. For example, it has been argued by the Google team~\cite{Boixo2018, Arute2019} that, if noise is local and incoherent, the output distribution of random quantum circuits can be well approximated by a global depolarizing (white noise) model, whose distribution is given by
\begin{align}
    p_{\rm noisy} (i) \approx q_U (i) = f p_U (i) + \frac{1-f}{N},
\label{eq:Google-model}
\end{align}
where $p_U (i) = \left|\bra{0\ldots 0} U \ket{i}\right|^2$ is the noiseless distribution 
and $f$ sets the noise strength; this output distribution has been supported by several theoretical studies 
\cite{Dalzell2021,Bouland2022,Deshpande2022}.

Given the probabilities of Eq.~(\ref{eq:Google-model}), the fluctuations of the Lorentz curves can be easily calculated and are given by
\begin{align}
    {\rm std}\,[F_q (k)] = f \, {\rm std}\,[F_p (k)],
\end{align}
where we used that $F_q(k) = f F_p(k) + (1-f) k/N$.
Thus, within this noise model, more realistic reference lines could be used by simply rescaling the cumulant fluctuations.  

To conclude, our results motivate further work to investigate how this technique performs on experimental implementation. We believe this indicator may serve as an alternative or complement other benchmarking techniques. However, in its present form, its applicability is limited by the number of measurements required, as it scales exponentially with the number of qubits. Nonetheless, we believe it might be possible to develop more efficient detection strategies, which shall be discussed elsewhere, once the connection between majorization and the potential to reach quantum advantage becomes clearer.


\acknowledgements

We acknowledge the Latin America Quantum Computer Center (LAQCC SENAI CIMATEC) for providing us access to their Quantum Simulator QLM ``KUATOMU'' (classical supercomputer). This work is supported in part by the National Council for Scientific and Technological Development, CNPq Brazil (Grant No. 409611/2022-0, and PCI-CBPF), CAPES, and it is part of the Brazilian National Institute for Quantum Information.

\bibliography{MajorizationRefs}

\end{document}